\documentclass[
aps,%
12pt,%
final,%
notitlepage,%
oneside,%
onecolumn,%
nobibnotes,%
nofootinbib,%
superscriptaddress,%
noshowpacs,%
centertags]%
{revtex4}

\usepackage{graphicx}

\def\refitem #1 {}

\begin{document}
\bibliographystyle{maik}

\title{Spontaneous large scale momentum exchange by microinstabilities: an analogy between tokamaks and space plasmas}

\author{\firstname{Giovanni}~\surname{Lapenta}}
\email{giovanni.lapenta@wis.kuleuven.be}

\affiliation{Centrum voor Plasma-Astrofysica, Departement Wiskunde,
Katholieke Universiteit Leuven, Celestijnenlaan 200B, 3001
Leuven, Belgium}

\affiliation{Plasma Theory Group, Los Alamos National Laboratory,
Los Alamos, New Mexico, USA}%

\date{\today}

\begin{abstract}
Based on a recent theory (Coppi, Nuclear Fusion, 42, 1, 2002) of spontaneous toroidal rotation in tokamaks (Lee et al, Phys Rev Lett,  91, 205003, 2003) and in astrophysical accretion disks, we propose that an analogous process could be at play also in the Earth space environment. We use fully kinetic PIC simulations to study the evolution of drift instabilities and we show that indeed a macroscopic velocity shear is generated spontaneously in the plasma. As in tokamaks, the microscopic fluctuations remain limited to the edge of the plasma channel but the momentum  spreads over the whole macroscopic system. 
\end{abstract}

\maketitle

\section{Introduction}

The understanding of the interplay between microinstabilities and macroscopic system-scale processes poses  a formidable challenge to current plasma physics research. The present paper attempts to attract the attention of the plasma physics research community to a possible analogy between space plasmas and tokamaks in the hope that a fruitful interdisciplinary exchange can lead to a new a deeper understanding. 

Among the small scale kinetic processes, the drift instabilities driven by density gradients have attracted considerable attention~\cite{huba}. It has long been known that the small scale electric field and magnetic field fluctuations induced by such waves can act as an effective collision by altering the momentum  of plasma particles in a manner analogous to regular classical collisions~\cite{gary80}. Such processes have been termed anomalous in the sense that when modeled within the framework of classical transport or fluid frameworks appear to lead to a much higher collisionality than is actually present~\cite{sagdeev}. 
But such an approach focuses on the small scale momentum exchanges and sees the microinstabilities as acting in a localized fashion akin to collisions. Instead, a number of recent findings suggest that drift instabilities can act to alter momentum on a global macroscopic scale. 

Recent tokamak research has incurred in a new 
crucial experimental finding: the spontaneous creation of toroidal rotation. In a number of experiments (including Alcator C-Mod, DIII-D, TCV, Tore-Supra, JET, TFTR, COMPASS) the plasma has been observed to acquire a toroidal rotation profile spontaneously in shots where no external input of toroidal momentum was imposed (see ref.~\cite{RiceJE:Obsamt,tcv,d3d} and references therein). Researchers at Alcator C-Mod have investigated the issue in depth~\cite{LeeWD:ObsAMT, RiceJE:Torram, RiceJE:Obsamt}. We refer the reader to their work for details and we limit the present discussion to highlight the findings most relevant to the present work. First, the generation of toroidal rotation  is spontaneous and  no external injection of angular momentum could be identified. Second, the  source of angular momentum appears to be located in the edge zone where angular momentum is observed to form first. Later during the shot, toroidal angular momentum is transported inward at a pace faster than predicted by neoclassical diffusivity. 

To explain the observed spontaneous angular rotation in tokamaks, a number of possibilities have been proposed. We refer the reader again to the existing literature from the Alcator C-Mod team for a discussion of all possibilities~\cite{RiceJE:Torram} and we limit the scope here to the particularly promising theoretical model proposed by Coppi~\cite{CoppiB}. Coppi noted an intriguing analogy between the spontaneous rotation observed in tokamak experiments and angular momentum exchange in accretion disks in astrophysical systems (such as forming stars or accreting black holes). The same core mechanism is at play in both systems: the exchange of angular momentum from the action of microscopic drift instabilities caused by the presence of pressure gradients. When applied to tokamak plasmas, Coppi's model identifies the source of angular momentum at the edge of the plasma column where fluctuations due to edge-localized pressure-driven modes scatter particles out of confinement and transfer angular momentum to the wall, thereby producing a net momentum in the plasma by reaction. The momentum created at the plasma edge is then transported inward and leads to the observed profiles of toroidal angular velocity measured in Alcator C-Mod. 

In the present work we revisit the model proposed by Coppi and extend the analogy to a different class of plasmas: the space plasmas present in the near Earth environment. The usefulness of this analogy is that space plasmas are amenable to direct satellite observation of small scale processes and a wealth of data is available  from past, current and planned missions.
The paper is organized as follows. In section 2, we summarize the processes leading to the spontaneous formation of velocity shear in space plasmas. In Section 3 we present our simulation approach to the problem and in Section 4 we present the results of our simulation study and compare them with satellite observations. Conclusions are drawn in Section 5.

\section{Large scale velocity shear formation  in space plasmas}
The spontaneous  formation of large scale velocity profiles has been recently discovered in space plasmas near the Earth~\cite{lapenta02,daughton02,lapenta03}. The Earth is surrounded by a magnetic field that in its interaction with the solar wind forms a complex and dynamical structure, called magnetosphere~\cite{kallenrode}. In the direction opposite to the Sun, approximately on the equatorial plane, the magnetic field is stretched to form the magnetotail where a current sheet is present and the magnetic field reverses sign. The northern part of the current sheet is magnetically linked to the north pole while the southern part is linked to the South pole. The current sheet covers a wide angular span in the night side of the Earth. Using the  Geocentric Solar Magnetospheric (GSM) coordinates, the toroidal direction moving around the Earth in the magnetic equatorial plane is called the dawn-dusk direction, the vertical direction along the magnetic poles is the north-south direction and the third direction is from the Sun towards the Earth. 

Recent studies of the current layer in the magnetotail have uncovered a previously unexpected result\cite{lapenta02,daughton02}: microinstabilities on the scale of the electron gyroscale (of the order of the Km) developing in the edge of the current layer exchange momentum with the plasma in the dawn-dusk direction not only in microscopic regions, but extending over the whole current layer, spanning distances that equal tens of thousands of Km.

The existence of density gradient driven instabilities in the lower-hybrid range (lower-hybrid drift instability, LHDI) has been an established fact for a long time, confirmed by numerous missions dating back from the early times of space exploration~\cite{lhdisat}. But it was always believed that such instabilities would affect the overall system in the form of substitute collisions leading to anomalous transport effects~\cite{gary80,huba}. The fact is, instead, that they lead directly  to plasma-waves exchanges of momentum on the macroscopic scale of the whole current layer. 

The analogy with the tokamak experiments summarized above is obvious: in both cases the toroidal (or dawn-dusk) flow is generated spontaneously in the edge of the the plasma but its effects are macroscopic leading to a toroidal flow throughout the plasma column. Furthermore, according to Coppi's model for the tokamak case and our model for the magnetotail, in both cases the cause is the presence of  fluctuations due to pressure-gradient driven modes.

However, we should warn the reader of some key differences that need to be considered in pursuing the analogy further. In geospace there is a toroidal magnetic  field but it is small or comparable with the poloidal field. Furthermore, the plasma thickness, once rescaled with  the ion inertial length,  is  smaller in the magnetotail (despite its physical size). The plasma beta is also different, being largely above unity in a large portion of the magnetotail. For these reasons the nature of the microinstabilities might be different and the analogy might not be complete.

\section{Modeling approach}
To simulate the effect of small scale drift instabilities on the macroscopic momentum profile, we consider an idealization of the conditions typical of the current sheet in the Earth magnetotail.  
The initial state is a Harris
sheet~\citep{harris}, with magnetic field and plasma density given
in GSM coordinates ($x$ aligned with the main tail
magnetic field in the Sun-Earth direction, $y$ aligned with the
current in the dawn-dusk direction, $z$ aligned with the main
gradients in the north-south direction) by
$
B_x = B_0 \tanh (z/L) $, 
$n =n_0 {\rm sech}^2 (z/L)$.

The current sheet half-width is chosen as $L =  d_i/2$, where $d_i =
c/\omega_{pi}$ is the ion inertia length. The temperature is uniform and a temperature ratio $T_i/T_e = 5$ and a
mass ratio $m_i/m_e=180$ are assumed. To consider drift instabilities, we conduct the simulation in the   2D $(y,z)$ domain
delimited by $-6.4 < y/L < 6.4$, $0 < z/L < 25.6$. 

To conduct the simulations, we rely on our long standing kinetic
code CELESTE3D (see Ref.~\cite{lapenta06} and references therein).
The code is 3D but for the present study the $x$ direction is not
used.  A grid with 128x256 cells is used with $\omega_{pi}\Delta t=0.1$.
In the $y$ direction, periodic boundary conditions are used.
In the  $z$ direction, the fields are subject to a Dirichlet
boundary condition. Note that the latter condition allow the exchange of momentum between the plasma and the boundary (that summarizes the interaction with the Sun-Earth system via the solar wind)~\cite{lapenta02,daughton02}.

\section{Simulation  results}
In the edge of the magnetotail current layer, the LHDI develops with properties described by previous theoretical studies~\cite{gary,daughton03}. Its free energy is provided by the diamagnetic drift
caused by the density gradient. The frequency of the LHDI is in the lower-hybrid range: 
$\Omega_{ce} > \omega > \Omega_{ci}$. The LHDI is a microinstability with peak growth at scales of the order of the electron gyroradius: $k_\perp \rho_e \sim 1$ and propagating in the direction normal to the magnetic field: $k_{||}=0$. While the fastest modes are on the electron gyroradius scale and are mostly electrostatic, longer wavelength ($k_\perp \sim \sqrt{\rho_e \rho_i}$) electromagnetic modes exist~\cite{winske,daughton03}. The LHDI remains mostly localized in the edge of the plasma because it is stabilized by the high beta in the centre of the layer~\cite{DrakeJF}.  

\begin{figure}
\begin{tabular}{c}
\includegraphics[width=8cm,height=3cm]{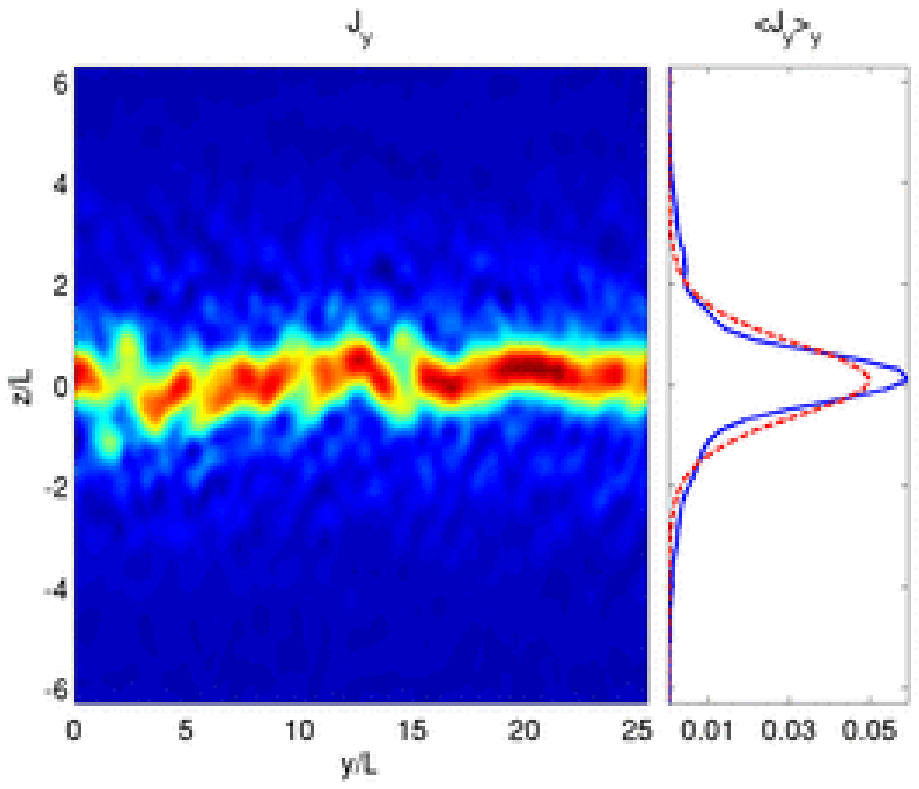}\\
\includegraphics[width=8cm,height=3cm]{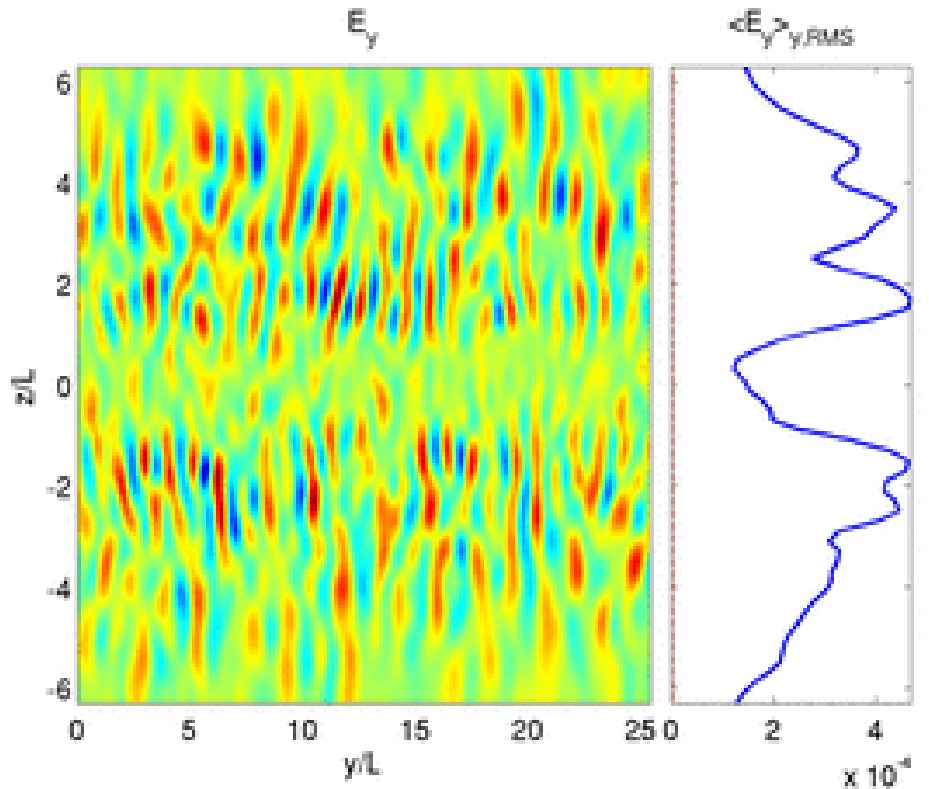}\\
\includegraphics[width=8cm,height=3cm]{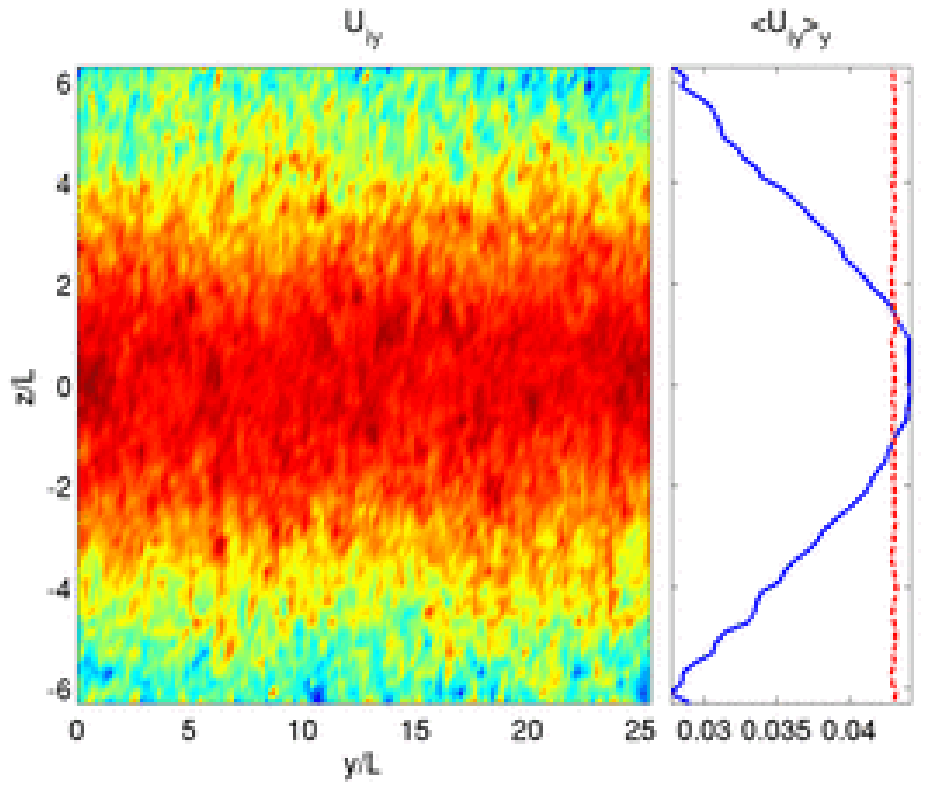}
\end{tabular}
\caption{Simulation of the evolution of the LHDI, time shown: $\omega_{ci}t=14.7$. From top to bottom: toroidal current, electric field and ion flow speed. On the right, the quantities averaged in $y$ are shown (solid blue line) compared with the initial profile (dashed red line). For the fluctuation toroidal electric field, the root mean square is shown.}
\label{fig1}
\end{figure}

Figure~\ref{fig1} shows the dawn-dusk (toroidal) current, electric field and velocity at time $\Omega_{ci} t=14.7$ when the LHDI is fully saturated. For each quantity, the left panel shows the 2D false color representation and the right panel reports the $y$-averaged profile (with the exception of the electric field fluctuation where instead of the mean, the RMS is shown).  The presence of the LHDI is highlighted by the recurring stripes of red and blue in the toroidal electric field. The fluctuations peak at around one ion inertial length (corresponding to $z=\pm 2L$) off the center and remain localized in the edges, away from the center of the current sheet. 

The non-linear effect of the LHDI is to cause the spontaneous formation of a velocity profile in the dawn-dusk direction. The bottom panel of Fig.~\ref{fig1} shows the  ion velocity at time $\Omega_{ci} t=14.7$: the velocity presents a small scale noise due to the presence of the LHDI and even of Langmuir waves but on the whole it has developed a large scale structure absent in the original plasma. The shear is evident in the  $y$-averaged profile, where the initial flat ion velocity profile is shown for reference. The saturation of the LHDI has caused a large exchange of momentum  that has altered the profile producing a large shear. While the instability is limited to the edge, the momentum exchange does not remain localized in the region where its cause is present but extends over the whole layer, including its center. 

The dawn-dusk current is also affected by the LHDI~\cite{lapenta02,lapenta03, ricci-lhdi,ricci-invited}: it is intensified and its initially straight surface is kinked. The kinking of the current layer has two separate scales with different causes. The smaller scale is still longer wavelength ($k_\perp \sim \sqrt{\rho_e \rho_i}$) than the bulk of the LHDI modes and it is due to the electromagnetic branch of the LHDI.   Instead, the longer wavelength kinking  is due to a secondary Kelvin-Helmholtz instability created by  the formation of the velocity shear caused by the LHDI~\cite{lapenta02}.

\begin{figure}
\includegraphics[width=8cm,height=8cm]{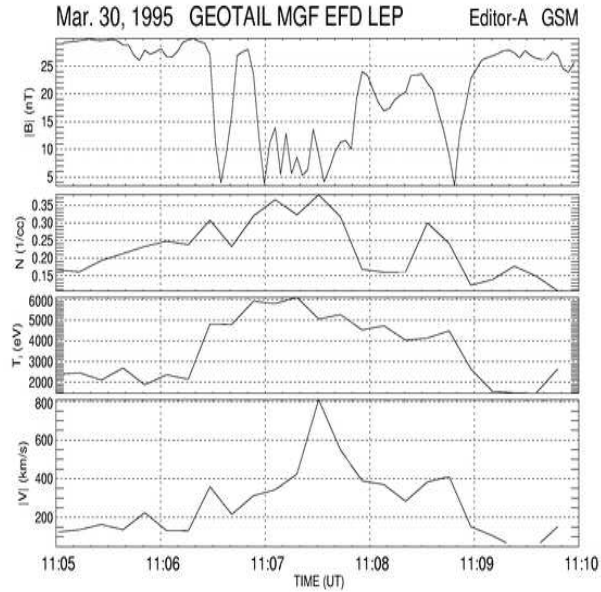} 
\caption{GEOTAIL magnetic field data and plasma data provided by T. Nagai, H. Hayakawa and T. Mukai through DARTS at the Institute of Space and Astronautical Science (ISAS) in Japan. Data for March 30, 1995 when Geotail was at a distance of $15.2 R_E$ along the Earth magnetotail \cite{ShinoharaI:Loweto}.}
\label{fig3}
\end{figure}

Evidence can be found in existing satellite databases to validate our results summarized above. As an example, Fig.~\ref{fig3} reports previously published results from the Geotail mission~\cite{ShinoharaI:Loweto}. The time interval corresponds to one instance of the satellite entering the magnetotail current layer where the magnetic field reverses (at about time 11:07UT), visible in Fig.~\ref{fig3} when the magnetic field amplitude drops to near zero.  As the satellite enters the layer, the plasma density increases, reaching its peak in the center and the temperature remains essentially constant within the layer. Therefore, the Harris equilibrium used in the simulations is a reasonable approximation of the real layer. The bottom panel shows the plasma speed, revealing the presence of a velocity shear with the peak-speed reached in the center and decreasing towards the edge. This finding is in remarkable agreement with the simulation results shown in the bottom panel of Fig.~\ref{fig1}. Additional Geotail data for the same crossing shows a strong presence of electrostatic fluctuations in the edge of the plasma~\cite{ShinoharaI:Loweto}, further supporting the scenario described above.

\begin{figure}
\includegraphics[width=8cm,height=3cm]{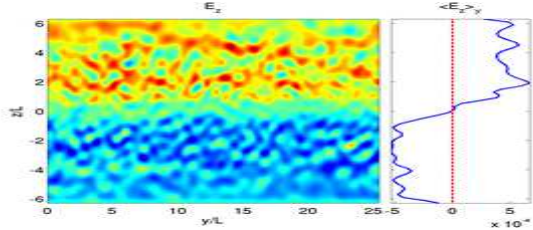} 
\caption{Vertical electric field for the same run as in fig.~\ref{fig1}, time shown: $\omega_{ci}t=14.7$. On the left, the field is shown averaged in $y$.}
\label{fig4}
\end{figure}

A complete theory of the precise mechanism allowing the small scale LHDI developing in the edge to create a large velocity shear will require future investigations. However, one key aspect of it has been revealed in a recent simulation investigation~\cite{daughton-prl}, highlighting the role of electrostatic fields. The formation of a velocity profile is accompanied by the formation of a large scale electrostatic field.  Figure \ref{fig4} shows for the same case in Fig.~\ref{fig1} the vertical component of the electric field. The saturation of the LHDI has created a large scale electric field that is primarily electrostatic in nature: the electric field shown is essentially curl-free and it is caused by charge separation~\cite{daughton-prl}. An obvious interpretation is that the small scale convective cells caused by the LHDI self-organize to form a large scale shear profile in a manner typical of many flow processes~\cite{finn-shear}.

\section{Conclusions}

We have reconsidered the recent theory by Coppi~\cite{CoppiB} relative to the analogy between the exchange of angular momentum in accretion disks and in tokamaks where a spontaneous toroidal rotation has been observed (e.g. Alcator C-Mod, DIII-D, TCV,Tore-Supra, JET, TFTR, COMPASS)~\cite{LeeWD:ObsAMT, RiceJE:Torram, RiceJE:Obsamt,tcv,d3d}. We have proposed above that a similar analogy can be extended to a new process recently discovered with a simulation  investigation~\cite{lapenta02,daughton02,lapenta03}: the spontaneous formation  of a velocity shear in the dawn-dusk direction in the Earth magnetotail by the action of drift waves (LHDI). We have outlined the possible analogy and we have presented a detailed simulation that investigates the conditions prevalent in the Earth environment. We have further showed satellite data that provide experimental evidence supporting the existence of the processes predicted by our simulations.
 
\begin{acknowledgments}
It is a pleasure to thank B. Coppi for the engaging conversations on his theory of spontaneous toroidal rotation that inspired the present paper. The author is also grateful for the hospitality of the PSFC at MIT and for the discussions with the staff there on the topic of spontaneous rotation in Alcator C-Mod. 
The present work is supported by the {\it Onderzoekfonds K.U. Leuven} (Research Fund KU Leuven), by the European Commission through the SOLAIRE network (MRTN-CT-2006-035484), by the NASA Sun Earth Connection Theory Program and by
 the LDRD program at the
Los Alamos National Laboratory. Work performed in part under the auspices of
the National Nuclear Security Administration of the U.S. Department
of Energy by the Los Alamos National Laboratory, operated by Los
Alamos National Security LLC under contract DE-AC52-06NA25396. The GEOTAIL magnetic field data and plasma data was provided by T. Nagai, H. Hayakawa and T. Mukai through DARTS at the Institute of Space and Astronautical Science (ISAS) in Japan. Simulations conducted in part on the
HPC cluster VIC of the Katholieke Universiteit Leuven.
\end{acknowledgments}

\end{document}